\documentclass[aip,amsmath,amssymb,
reprint,
]{revtex4-1}

\usepackage[caption = false]{subfig}
\usepackage{graphicx}
\usepackage{dcolumn}
\usepackage{bm}

\usepackage[utf8]{inputenc}
\usepackage[T1]{fontenc}
\usepackage{mathptmx}
\usepackage{etoolbox}
\usepackage{siunitx}
\DeclareSIUnit\angstrom{\text {Å}}
\usepackage{mathtools}
\usepackage{longtable}
\usepackage{xr}

\makeatletter
\def\@email#1#2{%
 \endgroup
 \patchcmd{\titleblock@produce}
  {\frontmatter@RRAPformat}
  {\frontmatter@RRAPformat{\produce@RRAP{*#1\href{mailto:#2}{#2}}}\frontmatter@RRAPformat}
  {}{}
}%
\makeatother

\DeclareMathOperator{\Tr}{Tr}
\usepackage{xcolor}
\usepackage[normalem]{ulem}

\begin{document}

\preprint{AIP/123-QED}


\title{Approaching the basis-set limit of the dRPA correlation energy with explicitly correlated and Projector Augmented-wave methods}

\author{Moritz Humer}
 \email{moritz.humer@univie.ac.at}
\affiliation{University of Vienna, Faculty of Physics \& Computational Materials Physics \& Vienna Doctoral School in Physics,  Boltzmanngasse 5, A-1090 Vienna, Austria}
\author{Michael E. Harding}
\affiliation{Institut f\"{u}r Nanotechnologie, Karlsruher Institut f\"u{r} Technologie (KIT), Campus Nord, Postfach 3640, D-76021 Karlsruhe, Germany}
\author{Martin Schlipf}
\affiliation{VASP Software GmbH, Sensengasse 8, A-1090 Vienna, Austria}
\author{Amir Taheridehkordi}
\affiliation{University of Vienna, Faculty of Physics and Center for Computational Materials Science, Kolingasse 14-16, A-1090 Vienna, Austria}
\author{Zoran Sukurma}
\affiliation{University of Vienna, Faculty of Physics \& Computational Materials Physics \& Vienna Doctoral School in Physics,  Boltzmanngasse 5, A-1090 Vienna, Austria}
\author{Wim Klopper}
\affiliation{Institut f\"{u}r Nanotechnologie, Karlsruher Institut f\"u{r} Technologie (KIT), Campus Nord, Postfach 3640, D-76021 Karlsruhe, Germany}
\affiliation{Institut f\"{u}r Physikalische Chemie, Karlsruher Institut f\"u{r} Technologie (KIT), Campus S\"{u}d, Postfach 6980, D-76049 Karlsruhe, Germany}
\author{Georg Kresse}
\affiliation{VASP Software GmbH, Sensengasse 8, A-1090 Vienna, Austria}
\affiliation{University of Vienna, Faculty of Physics and Center for Computational Materials Science, Kolingasse 14-16, A-1090 Vienna, Austria}

\date{\today}

\begin{abstract}
The direct random-phase approximation (dRPA) is used to calculate and compare atomization energies for the HEAT set and 10 selected molecules of the G2-1 set using both plane waves and Gaussian-type orbitals. We describe detailed procedures to obtain highly accurate and well converged results for the projector augmented-wave (PAW) method as implemented in the Vienna Ab-initio Simulation Package (VASP) as well as the explicitly correlated dRPA-F12 method as implemented in the TURBOMOLE package. The two approaches agree within chemical accuracy (\SI{1}{\kilo cal \per mol}) for the atomization energies of all considered molecules, both for the exact exchange as well as for the dRPA. The root mean-square deviation is \SI{0.41}{\kilo cal\per\mol} for the exact exchange (evaluated using density functional theory orbitals) and  \SI{0.33}{\kilo cal\per\mol} for exact exchange plus the random-phase approximation. 
\end{abstract}

\maketitle

\section{\label{Introduction}Introduction}
Density functional theory (DFT) is without question one of the most versatile approaches to approximate the many-body Schr\"odinger equation.\cite{Kohn1965} A great advantage is that calculations of the ground-state energy require only the calculation of the occupied manifold of the one-electron orbitals since unoccupied orbitals do not enter by construction in the Kohn-Sham (KS) energy functional. This makes the method not only fast but also makes it easy to construct suitable and compact basis sets for KS ground-state energy calculations. Similarly, reliable and transferable pseudopotentials can be constructed by choosing a small core radius and matching the scattering properties of atoms at a small set of energies. The advent of the projector augmented wave (PAW) method made the construction of potentials arguably even simpler, since the PAW method allows to reconstruct the all-electron orbitals, and hence by construction takes into account the exact shape of the wavefunctions.\cite{Bloechl1994,Kresse1999} Pseudopotentials, on the other hand, make approximations for the shape. Using the PAW method, constructing potentials for the entire periodic table is indeed now as simple as matching the scattering properties of atoms at a sufficiently small core radius, at a set of energies for sufficiently many angular quantum numbers. It has repeatedly been demonstrated that such PAW potentials yield excellent agreement with other all-electron methods. For instance, Paier and Kresse showed an in-depth comparison between the results for molecules, obtained from the PAW method and Gaussian-type orbitals,\cite{Paier2010} finding agreement within 1 kcal/mol. Studies comparing many different all-electron and pseudopotential codes also confirmed that the PAW method is capable of predicting reference type quality results for solids.\cite{lejaeghere2016} 

However, present density functionals are by no means infallible. In fact, in the last decades, we have seen an ever-increasing number of density functionals, addressing one or the other pitfall. It goes well beyond the scope of this introduction to attempt an even
cursory review, but, the most important recent developments include machine-learned (ML) density functionals that are trained on highly accurate reference data. Attempts to construct ML density functionals have been particularly successful for small molecules where fairly large training data sets are available or can be computed with reasonable effort.\cite{Bogojeski2020,Li2021,Nagai2020,Margraf2021,Dick20220,Kirkpatrick2021} This underlines the need for highly accurate reference data, in particular for condensed matter systems, where such calculations are largely missing.

Performing reference type calculations for solids is still an extraordinarily difficult task, for the following reasons. First, correlated wavefunction methods, which are penultimately required to accurately solve the many-body Schr\"odinger equation, are traditionally implemented in Gaussian-type orbital codes. Now, unfortunately, many of the standard correlation consistent basis sets\cite{Dunning1989} can not be used for solid-state calculations, since these basis sets involve diffuse orbitals leading to almost singular overlap matrices that can not be inverted. Thus, reaching basis-set convergence is not as simple as for molecular systems. Second, many of the correlated wavefunction methods have adverse scaling both concerning the number of occupied states as well as the basis-set size. For instance, the coupled-cluster singles doubles with perturbative triples method\cite{raghavachari_chemphyslett_157_479_1989} (CCSD(T))  scales with the 7$^{\rm{th}}$ power of the size of the molecular system (number of electrons and orbitals). It is hence far from trivial to perform converged all-electron calculations for solid-state systems, even if the basis-set problem had been solved. Finally, a method implemented for molecules might not be readily available for calculations using periodic boundary conditions. Often the four-center Coulomb integrals become very long ranged and storing all of them might not be possible for solids. Furthermore, special attention is required in dealing with the tail of the $1/r$ Coulomb potential, which, even in the simplest approximation (the random phase approximation), leads to long-range dipole interactions that fall off like $1/r^6$.  Hence, simple truncations, or hierarchical methods that treat long-range interactions using a lower-level theory are potentially unreliable.\cite{Yasmine2017,Yasmine2021}

Ultimately, plane waves are much simpler to use for solids and can be easily made complete by increasing the number of plane waves. But as already eluded to above, plane waves necessarily need to be used in unison with pseudopotentials or other types of projection methods, such as the PAW method. Now it might seem that one could apply pseudopotentials constructed for ground-state KS calculations also in correlated wavefunction calculations. Unfortunately, this hope turns out to be unjustified. For instance, Klimes, Kaltak and Kresse have demonstrated that the pseudized orbitals should match closely in their norm to the all-electron orbitals; an aspect usually not considered when PAW potentials are constructed.\cite{Klimes2014} If they do not match, the energy contributions related to high energy dipole fluctuations can be underestimated. Also, as for Gaussian basis set codes, it is important to describe not only the ground-state orbitals accurately but the inter-electron cusp needs to be captured as well. There is no a priori telling whether this will be the case for standard PAW potentials, although, in principle, plane waves should be sufficiently flexible to model the cusp. Therefore, validation of the PAW potentials against accurate reference calculations is very desirable.

Ideally, we would like to do this validation for a computationally "simple" method that allows one to reach accurate convergence using Gaussian basis sets and plane waves. Specifically, convergence with box size and plane-wave cutoff must be attainable. The validation method that we have chosen in the present work is the random-phase approximation (RPA) to the correlation energy. 
Bohm and Pines\cite{Bohm1951,Pines1952,Bohm1953} developed the RPA in a series of papers between 1951 and 1953. In 1957 Gell-Mann and Brueckner\cite{Gell-Mann1957} pointed out that the RPA correlation energy corresponds to a partial summation of the many-body perturbation theory expansion. Combining the adiabatic connection framework\cite{Langreth1975,Langreth1977,Gunnarsson1976} with the fluctuation-dissipation theorem\cite{Callen1951} also recovers the RPA as a low-order approximation. 
For simplicity, we use the RPA in a post-processing manner on top of Kohn-Sham Perdew-Burke-Ernzerhof\cite{Perdew1996} (PBE) orbitals. This not only simplifies the setup but the PBE functional is also consistently implemented in most codes, including the codes in the present study: VASP and TURBOMOLE.

Although the RPA is not an exceedingly accurate theory for the correlation energy it exposes many of the problems also present in more involved methods. For instance, Scuseria \textit{et al}\cite{Scuseria2008} established a connection between the RPA and coupled-cluster theory (direct ring coupled-cluster doubles theory). 
Furthermore, van-der-Waals-(vdW)-like interactions between fluctuating dynamic dipoles are fully accounted for by the RPA. These kinds of high energy fluctuations are potentially difficult to capture using pseudopotentials.\cite{Klimes2014} From a computational point of view, we note that the RPA is implemented in many different plane wave, atomic orbital, and Gaussian-type orbital codes:
Furche\cite{Furche2001} was the first to apply the RPA to molecules in a post-Kohn-Sham context. This marks the beginning of a renaissance of the RPA in quantum chemistry with a focus on improved lower scaling RPA algorithms.\cite{Furche2008,Kaltak2014} In the following years further studies, applying the RPA to different molecular systems including the HEAT\cite{Tajti2004} and G2-I AE\cite{Curtiss1991,Curtiss1998} sets, were published using plane waves,\cite{Fuchs2002,Harl2008,Olsen2013,Fuchs2005} Gaussian-type orbitals,\cite{Furche2001,Furche2008,Bates2013} or numeric atom-centered-orbital basis sets.\cite{Blum2009,Ren2012} A somewhat troublesome observation is that the atomization energies reported in these studies markedly disagree even for simple diatomics: for example, the N$_2$ RPA atomization energy was reported to be \SI{224}, \SI{223} and \SI{220}{\kilo cal\per\mol} by Olsen and Thygesen,\cite{Olsen2013} Furche and Van Voorhis\cite{Furche2005} and Ren~\textit{et al.}\cite{Ren2012}, respectively. This calls for a careful evaluation and publication of reference type results that can help to validate other codes and other pseudopotentials. Our ultimate goal is, of course, to use the validated PAW potentials for solid-state calculations and to apply them with methods beyond the RPA.\cite{Schaefer_2021,Booth2013,Ramberger2019,Grueneis2015,Zhang2019,Brandenburg2019,Gruber2018} This implies that we seek to reach chemical precision, that is, at least 1 kcal/mol. We show conclusively that this goal can be reached.

The remainder of the paper is organized as follows. In Sec.~\ref{II} the adiabatic-connection fluctuation-dissipation theory leading to the random-phase approximation is presented. The computational details are discussed in Sec.~\ref{III} followed by the results in Sec.~\ref{IV}. Finally, Sec.~\ref{V} is devoted to the conclusion.

\section{Theory}\label{II}

Hartree atomic units $\hbar = m_e = e = 1$ are used throughout this paper. The adiabatic-connection formalism\cite{Langreth1975,Langreth1977,Gunnarsson1976} links the non-interacting Hamiltonian adiabatically to the fully interacting Hamiltonian by introducing a coupling-strength parameter $\lambda$ that scales the electron-electron interaction at fixed electronic density $n(\bm{r})$. The exchange-correlation energy functional is then formally given as
\begin{equation}
    \centering
    E_{\rm xc}[n] = \frac{1}{2} \int\limits_0^1 d\lambda \iint d\bm{r} d\bm{r}' \frac{n(\bm{r})n^\lambda_{\rm xc}(\bm{r},\bm{r}')}{\lvert\bm{r}-\bm{r}'\rvert}~,
    \label{eq.2.a.1}
\end{equation}
where the exchange-correlation hole $n^\lambda_{xc}(\bm{r},\bm{r}')$ is linked to the response properties of the system 
by invoking the fluctuation-dissipation theorem\cite{Callen1951} (FDT)
\begin{equation}
    n^\lambda_{\rm xc}(\bm{r},\bm{r}') = -\frac{1}{\pi}\int\limits_0^\infty d\omega \frac{\chi^\lambda(\bm{r},\bm{r}',i\omega)}{n(\bm{r})}  - 
    \delta(\bm{r} - \bm{r}') ~.
    \label{eq.2.a.2}
\end{equation}
The linear response function $\chi^\lambda$ of the $\lambda$-interacting system fulfills a Dyson-type equation\cite{Gross1985,Petersilka1996}
\begin{align}
\begin{split}
    \centering
    \chi^\lambda(\bm{r},\bm{r}',i\omega) =& \chi^0(\bm{r},\bm{r}',i\omega)+ \iint d\bm{r}_1d\bm{r}_2 \chi^0(\bm{r},\bm{r}_1,i\omega) \\
    &\times \left(\frac{\lambda}{\lvert\bm{r}_1-\bm{r}_2\rvert} + f^\lambda_{\rm xc}(\bm{r}_1,\bm{r}_2,\omega)\right)\chi^\lambda(\bm{r}_2,\bm{r}',i\omega).
\end{split}
\label{eq.2.a.3}
\end{align}
Here $f_{\rm xc}$ denotes the exchange-correlation kernel at coupling strength $\lambda$ and $\chi^0(\bm{r},\bm{r}',i\omega)$ is the independent particle response function of the KS system
\begin{equation}
    \centering
    \chi^0(\bm{r},\bm{r'},i\omega) = \sum_{ij} (f_i - f_j)\frac{\phi^*_i(\bm{r})\phi_j(\bm{r})\phi^*_j(\bm{r}')\phi_i(\bm{r}')}{\epsilon_i - \epsilon_j - i\omega},
    \label{eq.2.a.4}
\end{equation}
with the KS orbitals $\phi_i$, their energies $\epsilon_i$ and occupation number $f_i$. Together, Eqs.~\eqref{eq.2.a.1}--\eqref{eq.2.a.4} define the exchange-correlation energy within the adiabatic-connection fluctuation-dissipation theory (ACFDT). These equations are exact if the KS ground-state density agrees with the true many-body ground-state density, however, the evaluation of the response function of the interacting system (Eq.~\eqref{eq.2.a.3}) requires some approximation for the exchange-correlation kernel. The simplest one---the RPA---is obtained by setting $f_{\rm xc}$ to zero.
This allows to write the exchange-correlation energy as $E^{\rm RPA}_{\rm xc} = E^{\rm HF}_{\rm x}[\{\phi^{\rm KS}\}] + E_{\rm c}^{\rm RPA}$. Here $E^{\rm HF}_{\rm x}$ is the Hartree-Fock (HF) exchange energy functional evaluated with KS orbitals $\phi^{\rm KS}$. Together with the kinetic energy and the Hartree term this corresponds to the HF energy functional evaluated using KS orbitals, which we will refer to as Hartree plus exact-exchange (HXX) energy throughout this paper. This name originates in DFT where usually only approximate exchange functionals are used. The direct RPA correlation energy $E^{\rm RPA}_{\rm c}$ is given by
\begin{equation}
    \centering
    E_{\rm c}^{\rm RPA} = \frac{1}{2\pi}\int\limits_0^\infty d\omega \Tr[ln(1-\chi^0(i\omega)\nu) + \chi^0(i\omega)\nu],
     \label{eq.2.a.5}
\end{equation}
with $\Tr[AB]\coloneqq \iint d\bm{r}d\bm{r}' A(\bm{r},\bm{r}')B(\bm{r},\bm{r}')$ and $\nu$ abbreviates the Coulomb kernel. 

\section{Computational details}\label{III}
The ACFDT-RPA atomization energies are assessed for the 26 molecules included in the HEAT set\cite{Tajti2004} and for 10 additional molecules (BeH, C$_2$H$_4$, C$_2$H$_6$, H$_2$CO, CH$_4$, CH$_3$OH, Li$_2$, LiH, LiF and N$_2$H$_4$) from the G2-1AE set.\cite{Curtiss1991,Curtiss1998} The resulting test set contains molecules involving first and second row atoms (\textit{i.e.}, H to F). The HEAT geometries are found in a paper by Harding {\textit et al}.\cite{Harding2008} while the G2-1AE geometries have been taken from the supplementary material of Ref.~\onlinecite{Haunschild2012}.

\subsection{Plane-wave basis-set calculations}
In a periodic code, an isolated atom
or molecule is simulated by placing it in a sufficiently large unit cell to suppress interactions between its periodic replica. To obtain reliable atomization energies, we kept the unit cell and basis-set size consistent between the atomic and molecular calculations. Symmetry broken unit cells were used to avoid fractional occupancies.
We use the Vienna Ab initio Simulation Package (VASP) that relies on the PAW method\cite{Bloechl1994,Kresse1999} and thus employs pseudopotentials. We found the set of PAW potentials listed in Table~\ref{tab.3.1} to yield the most accurate correlation energies. Note that the semi-core $1s$ states of Li and Be were treated as valence states.
The total ground-state energy in the RPA is given by the HF energy functional plus the ACFDT-RPA correlation energy. The RPA is employed in a post-processing manner and thus the necessary inputs --- the KS orbitals and eigenvalues --- were obtained using the GGA by Perdew, Burke, and Ernzerhof (PBE).\cite{Perdew1996} Note that the calculation of $E_{\rm c}^{\rm RPA}$ involves the evaluation of the independent particle response function and as such requires summation over many virtual orbitals (see Eq.~\eqref{eq.2.a.4}). The evaluation of $E_{\rm c}^{\rm RPA}$ implemented in VASP scales cubically with the system size\cite{Kaltak2014} and is thus expensive compared to the evaluation of $E^{\rm HF}[\{\phi^{\rm KS}\}]$. Hence, the HXX contribution to the total ACFDT energy was converged separately and subsequently added to the extrapolated RPA correlation energy. In the following, we discuss technical details concerning the HXX and ACFDT-RPA calculation.
By setting \texttt{LASPH\,=.TRUE.} in all DFT calculations, we take the aspherical contributions inside the PAW sphere exactly into account for the electrostatic energy as well as the exchange and correlation energy.

\begin{table}
\caption{\label{tab.3.1} List of the PAW potentials used in the present work. Shown are the orbitals that are treated as valence orbitals and the label of the respective pseudopotential in the VASP database. The radial cutoffs for each angular momentum quantum number are specified as $n\times r_{\rm cut}$, where $n$ specifies the number of projectors and $r_{\rm cut}$ is the radial cutoff in atomic units for the specific angular quantum number. The local potential corresponds to the all-electron potential which is replaced by a local pseudopotential below the radius $r_{\rm core}$ (a.u.).}
\begin{ruledtabular}
\begin{tabular}{llllllll}
atom & valence & label         & s   &  p  &   d   & $r_{\rm core}$ \\
\hline
H         & $1s$      & h     & 2$\times$ 0.8 & 1$\times$ 0.8 & -   & 0.7  \\
Li        & $1s2s2p$  & sv\_GW& 2$\times$ 1.2 \& 1$\times$1.3 & 2$\times$ 1.5 & 1$\times$ 1.5 & 1.0  \\
Be        & $1s2s2p$  & sv\_GW& 3$\times$ 1.1 & 2$\times$ 1.3 & 1$\times$ 1.3 & 1.0\\
C         & $2s2p$    & h\_GW & 3$\times$ 1.0 & 3$\times$ 1.1 & 2$\times$ 1.1 & 0.8 \\
N         & $2s2p$    & h\_GW & 3$\times$ 0.9 & 3$\times$ 1.1 & 2$\times$ 1.1 & 0.9 \\
O         & $2s2p$    & h\_GW & 3$\times$ 1.0 & 3$\times$ 1.1 & 2$\times$ 1.1 & 0.9 \\
F         & $2s2p$    & h\_GW & 2$\times$ 0.8 & 3$\times$ 1.05 & 2$\times$ 1.05 & 0.8 \\
\end{tabular}
\end{ruledtabular}
\end{table}

\subsubsection{HXX energy}\label{III.A}
In the evaluation of the Fock operator, the Coulomb kernel is given by
\begin{equation}
    \nu_{\bm{G},\bm{G}'}(\bm{q})=\frac{4\pi }{V \lvert \bm{G}+\bm{q} \rvert^2}\delta_{\bm{G},\bm{G}'}
    \label{eq.3.a.1}
\end{equation}
and is thus ill defined at $\bm{G} + \bm{q} = 0$, where $V$ is the volume of the unit cell, $\bm{q}$ are the crystal momentum points chosen to sample the Brillouin zone, and we denote the reciprocal lattice vectors with $\bm{G}$.  A similar divergence issue occurs for the Hartree term but by combining the Hartree term with the electron-ion and ion-ion term, a well defined expression for the total electrostatic energy can be obtained.\cite{Ihm1979} In order to deal with the $\bm{G} + \bm{q} = 0$ component in the exchange operator, the Coulomb kernel is truncated in real space\cite{Spencer2008} to a sphere with unit-cell volume $V$ and radius $R_c$. This yields the following modified Coulomb kernel 
\begin{align}
\begin{split}
    \centering
    &\nu_{\bm{G},\bm{G}'}(\bm{q})= \\
    &=\left\{ 
    \begin{array}{ll}
      \frac{4\pi }{V\lvert\bm{G} + \bm{q}\rvert^2}\left[1 - \rm{cos}(|\it{\bm{G}} + \bm{q}| R_c)\right]\delta_{\bm{G},\bm{G}'}  & ,|\bm{q} + \bm{G} | \ne 0 \\
      \frac{2\pi R_c^2}{V}\delta_{\bm{G},\bm{G}'} & ,|\bm{q} + \bm{G} | = 0.
    \end{array}
    \right. 
\end{split}
\end{align}
This effectively cuts off inter-molecular interactions in the exchange energy.
The HXX energy has been evaluated using \SI{24}{\angstrom} boxes and a \SI{1000}{\electronvolt} energy cutoff. The resulting atomization energies are converged to \SI{1}{\milli\electronvolt} indicating that inter-molecular static dipole-interactions are well suppressed by the large cell and thus no volume extrapolation is required.
For a detailed discussion on cell size extrapolation for isolated molecules in VASP, we also refer to the supplementary material of Ref.~\onlinecite{Schaefer2021}. 

\subsubsection{ACFDT-RPA correlation energy}\label{III.B}
Similar to the exchange operator, the RPA correlation energy, Eq.~\eqref{eq.2.a.5}, is also ill-defined at $\bm{G}~=~0$ due to the Coulomb kernel, Eq.~\eqref{eq.3.a.1}. Unfortunately, the truncation procedure outlined above is not applicable, since it results in an oscillatory convergence of the correlation energy related to the cosine modulations in the Coulomb kernel.
Hence, alternative procedures need to be found to deal with this issue.
In a standard RPA calculation in VASP, all components $\bm{G}+\bm{q} = 0$ are excluded in the calculation of the trace in Eq.~\eqref{eq.2.a.5}, that is all rows and columns for $\bm{G}+\bm{q} = 0$  are suppressed. This introduces an error that is proportional to the volume of the unit cell, since important long-range contributions related to the correlation energy are not included. We correct for this by calculating the
Taylor expansion of the polarizability $\chi^{0}_{\bm{G},\bm{G}'}(\bm{q},\omega)$  for $|\bm{q}|  \to 0$. The wings are obtained by setting $|\bm{G}'|=0$ or $|\bm{G}|=0$ and 
approaching $|\bm{q}|  \to 0$ from three different Cartesian directions. 
The head is a $3 \times 3$ tensor describing the polarizability of the molecule at $\bm{G}'=0$ and $\bm{G}=0$ for infinitesimal small wave vectors $|\bm{q}|$. For the three Cartesian directions, the head and wings are added back to the 
polarizability matrix and a wave vector dependent correlation energy (\textit{i.e.}, dependent on the direction of $\hat{\bm{q}}$) is calculated. The final correlation energy is the average of the values obtained for the three Cartesian directions.
This specific treatment is selected by specifying \texttt{LOPTICS=.TRUE.} during the RPA calculations in VASP, and it significantly improves the cell size convergence. Note that this correction has not always been used in the past for calculations on small molecules. If the correction is not applied, it is mandatory to correct for the resulting error, which is proportional to the inverse of the volume of the unit cell.\cite{Schaefer2021}
Even with this correction, a residual undesirable contribution that is proportional to $1/V^2$ prevails. This contribution relates to the vdW interaction between the repeated atoms or molecules (see second next paragraph).

The calculation of the independent particle response function $\chi^0$ requires a large number of virtual orbitals and their respective eigenvalues. To avoid any basis-set induced errors, all virtual orbitals spanned by the plane-wave basis set are calculated by exact diagonalization of the KS Hamiltonian, and hence as many orbitals as plane waves in the basis set are included in the calculation of $\chi^0$. Note that only plane waves with a kinetic energy smaller than $E_{\rm cut}$ are included in the basis set.
The RPA correlation energy converges slowly with respect to the dimension of the response function $\chi^0$ governed by the energy cutoff $E^\chi_{cut}$. Specifically, the response function includes only components up to 

\begin{equation}
    \centering
    \frac{\hbar^2 | \bm{q}+\bm{G}| ^2  }{2 m_e} < E^\chi_{cut}.
\end{equation} 

For sufficiently large values of $E^\chi_{cut}$, the correlation energy shows the following functional behaviour\cite{Harl2008}
\begin{equation}
    \centering
    E_{\rm c}(E^\chi_{\rm cut}) = E^\infty_{\rm c} + \frac{A}{(E^\chi_{\rm cut})^{3/2}}~,
    \label{eq.3.b.1}
\end{equation}
allowing to extrapolate to infinite cutoffs. This procedure is done automatically for RPA calculations in VASP based on eight different cutoff values $E^\chi_{\rm cut}$. The key trick is to evaluate the independent particle response function for the largest value of these cutoffs $E^\chi_{\rm cut}$, then progressively remove rows and columns from the matrix and calculate the correlation energy for each of these truncated response function arrays.
If the largest cutoff $E^\chi_{\rm cut}$ equals two-thirds of the plane-wave cutoff for the orbitals, $E^\chi_{\rm cut} = 2/3 E_{\rm cut}$, then the automatic extrapolation is accurate to few meV.
Specifically, we systematically increased the plane-wave cutoff $E_{\rm cut}$
and $E^\chi_{\rm cut}= 2/3 E_{\rm cut}$ and found that the correlation energy contribution to the atomization energy is  converged to \SI{8}{\milli\electronvolt} (0.2 kcal/mol) for all molecules using an  \SI{800}{\electronvolt} plane-wave cutoff. For these convergence tests, we used $(\SI{8}{\angstrom})^3$ unit cells for both the atoms and molecules .

\begin{figure}%
\centering
\subfloat[LiH]{%
\includegraphics{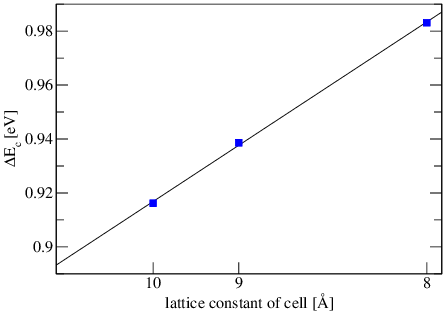}}\\ \vspace{-0.3cm}
\subfloat[N$_2$]{%
\includegraphics{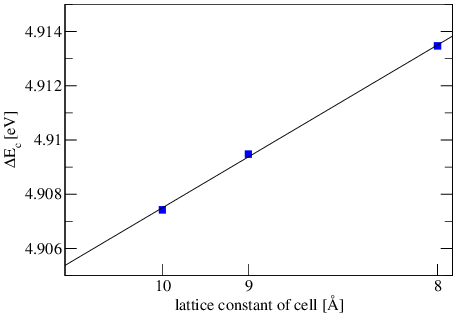}}
\caption{ Volume dependence of the correlation contribution to the atomization energy $\Delta E_{\rm c}$ of (a) LiH and (b) N$_2$. The calculated correlation energy (boxes) is extrapolated to the infinite cell size limit (solid line). The x-axis is given in a $V^{-2}$ scale. Note that the vertical axis scales differ in the top and bottom panels.}
\label{fig.3.b.1}
\end{figure}

As already eluded to above, the RPA correlation energy also needs to be converged with respect to the cell size. Note that, this is made easier in the present work since we only desire correlation energy differences. Fixing the energy cutoff now to  \SI{800}{\electronvolt} and  performing RPA calculations using ($\SI{8}{\angstrom})^3$, $(\SI{9}{\angstrom})^3$ and $(\SI{10}{\angstrom})^3$ cells we can estimate the errors related to the interaction between repeated images. The correlation contribution to the atomization energy $\Delta E_{\rm c}$ obeys a $V^{-2}$ volume dependence (see Fig.\,\ref{fig.3.b.1}) due to the dynamic dipole interactions between periodically repeated entities. These are taken into account by  extrapolating $\Delta E_{\rm c}$ to the infinite cell size limit using
\begin{equation}
    \centering
    \Delta E_{\rm c}(V) = \Delta E^\infty_{\rm c} + \frac{A}{V^{2}}~.
    \label{eq.3.b.2}
\end{equation}
In Fig.~\ref{fig.3.b.1}, the extrapolation of $\Delta E_{\rm c}$ is shown graphically for LiH and N$_2$. For instance, the LiH correlation energy difference (Fig.~\ref{fig.3.b.1} (a)) changes significantly between the $(\SI{10}{\angstrom})^3$ cell  (\SI{0.916}{\electronvolt}) and the infinite cell size limit (\SI{0.893}{\electronvolt}). Similar behavior is observed for the other lithium-compounds (Li$_2$ and LiF) as well as for BeH suggesting that, for $(\SI{10}{\angstrom})^3$ cells, there are still sizable long-range interactions present between the repeated molecules and atoms. On the other hand, N$_2$ (Fig.~\ref{fig.3.b.1} (b)) converges quickly with respect to the unit cell volume. Note the different scales of the y-axis in Fig.~\ref{fig.3.b.1}. Hence for N$_2$, only a minor difference of \SI{2}{\milli\electronvolt} is observed between the $($\SI{10}{\angstrom}$)^3$ cell result and the infinite cell size limit. For the majority of the molecules, long-range vdW interactions are sufficiently suppressed in the $($\SI{10}{\angstrom}$)^3$ cells resulting in converged correlation energy differences. However, for systems with spatially diffuse highest molecular orbitals such as the considered Li-compounds as well as BeH an extrapolation to the infinite cell size limit is necessary. In order to be consistent, all correlation energies were extrapolated to the infinite cell size limit using $($\SI{8}{\angstrom}$)^3$, $($\SI{9}{\angstrom}$)^3$ and $($\SI{10}{\angstrom}$)^3$ results at an \SI{800}{\electronvolt} plane-wave cutoff.

\subsection{Gaussian basis-set calculations}

The Gaussian basis-set calculations were carried out using the TURBOMOLE program package.\cite{TURBOMOLE} 
The ``direct'' RPA approach (dRPA)\cite{Klopper2011} was applied using converged PBE orbitals. The density-functional computations were performed with regular two-electron integrals and employing TURBOMOLE's quadrature grid 5a.\cite{C9CP02382H}
To ensure rapid convergence to the basis-set limit, the corresponding dRPA computations were carried out in terms of explicitly correlated direct-ring-coupled-cluster-doubles (drCCD-F12), which is equivalent to dRPA-F12 calculations.\cite{Hehn2013,Hehn2015}
In order to monitor the convergence to the basis-set limit, the correlation-consistent basis set series aug-cc-pCV$X$Z  ($X$= D, T, Q, 5, 6) \cite{Dunning1989,Kendall1992,Woon1995,Wilson1996} were employed in a deciontracted fashion in all computations for the elements H, C, N, O, and F. Due to the fact that decontraction of the aug-cc-pCV$X$Z basis leads to (numerical) linear dependencies, the $sp$-sets were replaced by those of the corresponding primitive $sp$ sets of the aug-cc-pV$X$Z \cite{Dunning1989,Kendall1992,Wilson1996} basis sets.

Besides the regular one-electron basis set, dRPA-F12 calculations require three auxiliary basis sets.
These are  
a basis set for the density-fitting approximation to certain two-electron integrals (MP2-fitting basis; CBAS),
a basis set for the density-fitting approximation to the matrix representation of the exchange operator of HF theory (exchange-fitting basis; JKBAS),
and a complementary auxiliary basis set (CABS) for the dRPA-F12 calculations. 
In order to minimize effects due to the incompleteness of the MP2-fitting basis, large MP2-fitting basis sets optimized for the aug-cc-pCV6Z basis were used in all computations for the elements H, C, N, O, and F. The optimization was done as described in Ref.~\onlinecite{Weigend1998}, and resulted in a 14s13p12d10f9g7h6i2k set for H and 22s21p20d19f18g16h15i8k sets for C, N, O, and F. 
Similar considerations led to an optimized large JKBAS 
for the elements H, C, N, O, and F, which resulted in 
13s12p5d4f1g,
24s18p13d12f12g9h5i,
27s18p15d14f10g7h8i,
and 
28s18p15d16f11g9h6i
for H, C, N, O, and F, respectively.
In the case of Li and Be, the JKBAS was taken from Ref.~\onlinecite{10.1002/jcc.20702}.

A restricted reference determinant was employed for closed-shell systems, all calculations for open-shell systems were performed using an unrestricted reference determinant. The HF self-consistent field (HF-SCF) energy was improved by taking into account single excitations into the CABS (CABS singles correction). The dRPA-F12 calculations were carried out using ansatz 2A [T+V] and the fixed amplitudes (sp) approximation with respect to spin-flipped geminals as proposed in
Ref.~\onlinecite{Tew2010}. In all cases, a Slater-type geminal with an exponent of $\gamma$ = 1.4 a$_{0}^{-1}$ was used. All convergence criteria have been set to at least 10$^{-11}$.

\section{Results}\label{IV}
In the following two subsections, we report the calculated non-relativistic HXX and RPA atomization energies for the 26 molecules included in the HEAT set and the 10 selected molecules of the G2-1 set. We compare plane-wave and Gaussian basis-set calculations and provide scalar relativistic results in the Supplementary material. The difference between the molecular energy $\epsilon(M)$ and the energies of the individual atoms $\epsilon(X)$ yields the atomization energy $D(M)$ of a molecule
\begin{equation}
    D(M) = \sum\limits_{X \in \mathrm{atoms}} n_X\epsilon(X) - \epsilon(M),
    \label{eq.4.0.1}
\end{equation}
where $n_X$ labels how many atoms of the species $X$ are included in the molecule.

\subsection{HXX atomization energies}
Table~\ref{tab.4.1} summarizes the HXX atomization energies obtained using plane waves (PW) and Gaussian-type orbitals (GTO). The maximal discrepancy between the two approaches is\break \SI{0.85}{\kilo cal\per\mol}, and the root-mean-square deviation amounts to \SI{0.41}{\kilo cal\per\mol}. We note that the plane-wave results are systematically smaller than the GTO calculations (\textit{i.e.}, underbinding in the PAW case). Fig.~\ref{fig.4.1} visualizes the difference between the atomization energies predicted by the two methods $\Delta E^{\rm HXX} = E^{\rm HXX}_{\rm PW} - E^{\rm HXX}_{\rm GTO}$. Paier \emph{et al.}\cite{Paier2010} observed a similar behavior when comparing PBE and PBE0 atomization energies for plane waves and GTOs. They suggested that these discrepancies originate either from basis-set superposition errors (BSSE) in the case of the GTO calculations or from the frozen-core approximation used in the PAW method. 

\LTcapwidth=0.745\textwidth
\begin{longtable*}{@{\extracolsep{\fill}}lrrr}
\caption{\label{tab.4.1}Hartree plus exact exchange (HXX) atomization energies using PBE orbitals for the HEAT test set\cite{Tajti2004} and selected molecules of the \hbox{G2-1} test set.\cite{Curtiss1991,Curtiss1998} The energies were calculated using plane-wave (PW) and Gaussian (GTO) basis sets both without relativistic corrections. Energies are given in \si{\kilo cal\per\mol}. The differences between PW and GTO are displayed in the column $\Delta_{\rm PW - GTO}$. Additionally, the root-mean-square deviation (RMSD) and mean signed deviation (MSD) are given in the final rows.} \\
\hline\hline
Molecule & \multicolumn{1}{c}{HXX$_{\rm PW}$} & \multicolumn{1}{c}{HXX$_{\rm GTO}$} & \multicolumn{1}{c}{$\Delta_{\rm PW - GTO}$} \\ \hline
\endfirsthead
\hline\hline
Molecule & \multicolumn{1}{c}{HXX$_{\rm PW}$} & \multicolumn{1}{c}{HXX$_{\rm GTO}$} & \multicolumn{1}{c}{$\Delta_{\rm PW - GTO}$} \\ \hline
\endhead
\hline
\hline
\endfoot
C$_2$H      & 176.15                              & 176.52                                 & -0.37                                              \\
C$_2$H$_2$     & 290.34                              & 290.71                                 & -0.37                                              \\
CF       & 72.33                               & 72.52                                  & -0.19                                              \\
CH       & 55.82                               & 56.05                                  & -0.23                                              \\
CH$_2$      & 154.06                              & 154.34                                 & -0.28                                              \\
CH$_3$      & 242.13                              & 242.54                                 & -0.41                                              \\
CN       & 71.20                               & 71.55                                  & -0.35                                              \\
CO       & 169.51                              & 169.63                                 & -0.12                                              \\
CO$_2$      & 233.84                              & 234.32                                 & -0.48                                              \\
F$_2$       & -42.60                              & -42.61                                 & 0.01                                               \\
H$_2$       & 83.92                               & 84.07                                  & -0.15                                              \\
H$_2$O      & 154.82                              & 155.17                                 & -0.35                                              \\
H$_2$O$_2$     & 128.96                              & 129.45                                 & -0.49                                              \\
HCN      & 194.54                              & 194.83                                 & -0.29                                              \\
HCO      & 175.50                              & 175.86                                 & -0.36                                              \\
HF       & 96.29                               & 96.42                                  & -0.13                                              \\
HNO      & 71.92                               & 72.47                                  & -0.55                                              \\
HO$_2$      & 58.76                               & 59.22                                  & -0.46                                              \\
N$_2$       & 110.60                              & 110.95                                 & -0.35                                              \\
NH       & 49.76                               & 49.99                                  & -0.23                                              \\
NH$_2$      & 116.02                              & 116.41                                 & -0.39                                              \\
NH$_3$      & 199.61                              & 200.09                                 & -0.48                                              \\
NO       & 46.40                               & 46.87                                  & -0.47                                              \\
O$_2$       & 25.43                               & 25.89                                  & -0.46                                              \\
OF       & -30.57                              & -30.38                                 & -0.19                                              \\
OH       & 67.39                               & 67.58                                  & -0.19                                              \\
BeH      & 49.19                               & 49.28                                  & -0.09                                              \\
C$_2$H$_4$     & 424.39                              & 425.01                                 & -0.62                                              \\
C$_2$H$_6$     & 547.57                              & 548.41                                 & -0.84                                              \\
CH$_3$OH     & 364.34                              & 365.07                                 & -0.73                                              \\
CH$_4$      & 326.74                              & 327.23                                 & -0.49                                              \\
H$_2$CO     & 251.10                              & 251.58                                 & -0.48                                              \\
Li$_2$      & 3.49                                & 3.48                                   & 0.01                                               \\
LiH      & 33.81                               & 33.81                                  & 0.00                                               \\
LiF      & 87.04                               & 87.10                                  & -0.06                                              \\
N$_2$H$_4$     & 262.80                              & 263.65                                 & -0.85                                   \\
\hline
RMSD    & & & 0.41  \\
MSD      & & & -0.35      \\
\end{longtable*}

In the present work, we have taken great care to remove any remaining uncertainties from the GTO results with respect to the basis-set size. Thus, the remaining uncertainties are well below  \SI{0.04}{\kilo cal \per\mol}.
To estimate the effect of long-range interactions between adjacent molecules on the plane-wave results we performed calculations for a series of unit cells, as explained before. We found that long-range effects are sufficiently suppressed by using (\SI{24}{\angstrom})$^3$ unit cells. Furthermore, the plane-wave results were also converged with respect to the energy cutoff to attain about the same accuracy of \SI{0.04}{\kilo cal\per\mol}. As such, we do not expect BSSE or finite-size errors to affect the reported numbers. 

Hence, we need to conclude that the remaining error is due to the PAW approximation.
Specifically, the core electrons are kept frozen in the PAW calculation at the level
of the PBE reference atom calculations. This is, in fact, a somewhat uncontrolled approximation, 
but one that is very difficult to remove in the PAW methodology. Specifically, the PAW approximation is so fast and numerically robust because the core electrons are kept frozen, which removes a huge energy contribution from the total energy. We speculate that this causes an underbinding because the core relaxation effects are likely to be larger for the lower symmetry molecules than for atoms: keeping the core rigid and frozen results in an upper bound for the total energy, possibly with a larger underestimation of the energy for molecules. We, furthermore, modified the energies at which the partial waves are constructed in the PAW method, but found no systematic way to improve the atomization energies. 

The assumption that freezing the core electrons is the remaining source of the error, is
also corroborated by the observation that H$_2$, LiH, Li$_2$ as well as BeH show a remarkably good agreement between both methods. In these cases, the $1s$ semi-core states are treated as valence states in the plane-wave calculations (and, as matter of fact, also in the GTO calculations). The remaining discrepancies are certainly within the error bars of the present calculations.

We conclude that the majority of the deviations between plane-wave and GTO calculations at the HXX level are due to the frozen-core approximation used in the PAW method. Hence, we estimate the residual error of the used GW PAW potentials (see Table~\ref{tab.3.1}) to be around \SI{0.5}{\kilo cal\per\mol} for Hartree-Fock type calculations. The agreement between both methods is certainly within the desired chemical precision (1 kcal/mol) and satisfactory for KS, Hartree-Fock, and most likely any other mean-field calculation. 

\subsection{RPA atomization energies}

We obtain the plane-wave total RPA atomization energies by adding the HXX atomization energies (see Table~\ref{tab.4.1}) and the RPA correlation energies (RPAc). Table~\ref{tab.4.2} summarizes the non-relativistic RPA results using plane waves as well as GTOs. We refer to the supplementary material for the scalar relativistic RPA atomization energies.
Comparing the plane-wave RPA atomization energies to the GTO results, we find the largest deviation of \SI{-0.8}{\kilo cal\per\mol} for H$_2$O$_2$ while the root-mean-square deviation amounts to \SI{0.33}{\kilo cal\per\mol}. The uncertainty of the GTO calculations with respect to the complete-basis-set limit is at most \SI{0.2}{\kilo cal\per\mol}. Regarding the plane-wave results, we estimate the errors due to the finite basis-set size and the extrapolation to infinite cell size to be around \SI{0.2}{\kilo cal\per\mol} as well.

\LTcapwidth=0.745\textwidth
\begin{longtable*}[H]{@{\extracolsep{\fill}}lrrr}
\caption{\label{tab.4.2}RPA atomization energies using PBE orbitals for the HEAT test set\cite{Tajti2004} and selected molecules of the G2-1 test set.\cite{Curtiss1991,Curtiss1998} The energies were calculated using plane-wave (PW) and Gaussian (GTO) basis sets without relativistic corrections. Energies are given in \si{\kilo cal\per\mol}. The differences between PW and GTO are displayed in the column $\Delta_{\rm PW - GTO}$. Additionally, the root-mean-square deviation (RMSD) and mean signed deviation (MSD) are given in the final rows.} \\

\hline\hline
Molecule & \multicolumn{1}{c}{RPA$_{\rm PW}$} & \multicolumn{1}{c}{RPA$_{\rm GTO}$} & \multicolumn{1}{c}{$\Delta_{\rm PW - GTO}$} \\ \hline
\endfirsthead
\hline\hline
Molecule & \multicolumn{1}{c}{RPA$_{\rm PW}$} & \multicolumn{1}{c}{RPA$_{\rm GTO}$} & \multicolumn{1}{c}{$\Delta_{\rm PW - GTO}$} \\ \hline
\endhead
\hline
\hline
\endfoot
C$_2$H      & 244.95                              & 244.52                                 & 0.43                                               \\
C$_2$H$_2$     & 381.78                              & 381.17                                 & 0.61                                               \\
CF       & 120.38                              & 120.58                                 & -0.20                                              \\
CH       & 81.19                               & 81.21                                  & -0.02                                              \\
CH$_2$      & 179.62                              & 179.59                                 & 0.03                                               \\
CH$_3$      & 294.51                              & 294.39                                 & 0.12                                               \\
CN       & 172.73                              & 172.32                                 & 0.41                                               \\
CO       & 244.39                              & 244.46                                 & -0.07                                              \\
CO$_2$      & 364.33                              & 364.76                                 & -0.43                                              \\
F$_2$       & 30.01                               & 30.56                                  & -0.55                                              \\
H$_2$       & 108.73                              & 108.69                                 & 0.04                                               \\
H$_2$O      & 222.82                              & 223.19                                 & -0.37                                              \\
H$_2$O$_2$     & 255.01                              & 255.81                                 & -0.80                                              \\
HCN      & 299.49                              & 298.87                                 & 0.62                                               \\
HCO      & 263.42                              & 263.65                                 & -0.23                                              \\
HF       & 132.51                              & 132.59                                 & -0.08                                              \\
HNO      & 198.93                              & 199.14                                 & -0.21                                              \\
HO$_2$      & 165.42                              & 166.00                                 & -0.58                                              \\
N$_2$       & 223.72                              & 223.34                                 & 0.38                                               \\
NH       & 82.47                               & 82.35                                  & 0.12                                               \\
NH$_2$      & 179.33                              & 179.17                                 & 0.16                                               \\
NH$_3$      & 290.64                              & 290.55                                 & 0.09                                               \\
NO       & 147.86                              & 147.93                                 & -0.07                                              \\
O$_2$       & 112.85                              & 113.30                                 & -0.45                                              \\
OF       & 46.19                               & 46.40                                  & -0.21                                              \\
OH       & 103.31                              & 103.39                                 & -0.08                                              \\
BeH      & 50.88                               & 50.57                                  & 0.31                                               \\
C$_2$H$_4$     & 537.72                              & 537.30                                 & 0.42                                               \\
C$_2$H$_6$     & 683.88                              & 683.68                                 & 0.20                                               \\
H$_2$CO     & 355.46                              & 355.69                                 & -0.23                                              \\
CH$_4$      & 404.94                              & 404.77                                 & 0.17                                               \\
CH$_3$OH     & 490.36                              & 490.83                                 & -0.47                                              \\
Li$_2$      & 18.88                               & 18.91                                  & -0.03                                              \\
LiH      & 54.41                               & 54.48                                  & -0.07                                              \\
LiF      & 127.35                              & 127.20                                 & 0.15                                               \\
N$_2$H$_4$     & 426.59                              & 426.63  
    & -0.04                                                 \\
    \hline
RMSD  &   &   & 0.33 \\
MSD     &   &   & -0.03
\end{longtable*}

\begin{figure*}
    \centering
    \includegraphics{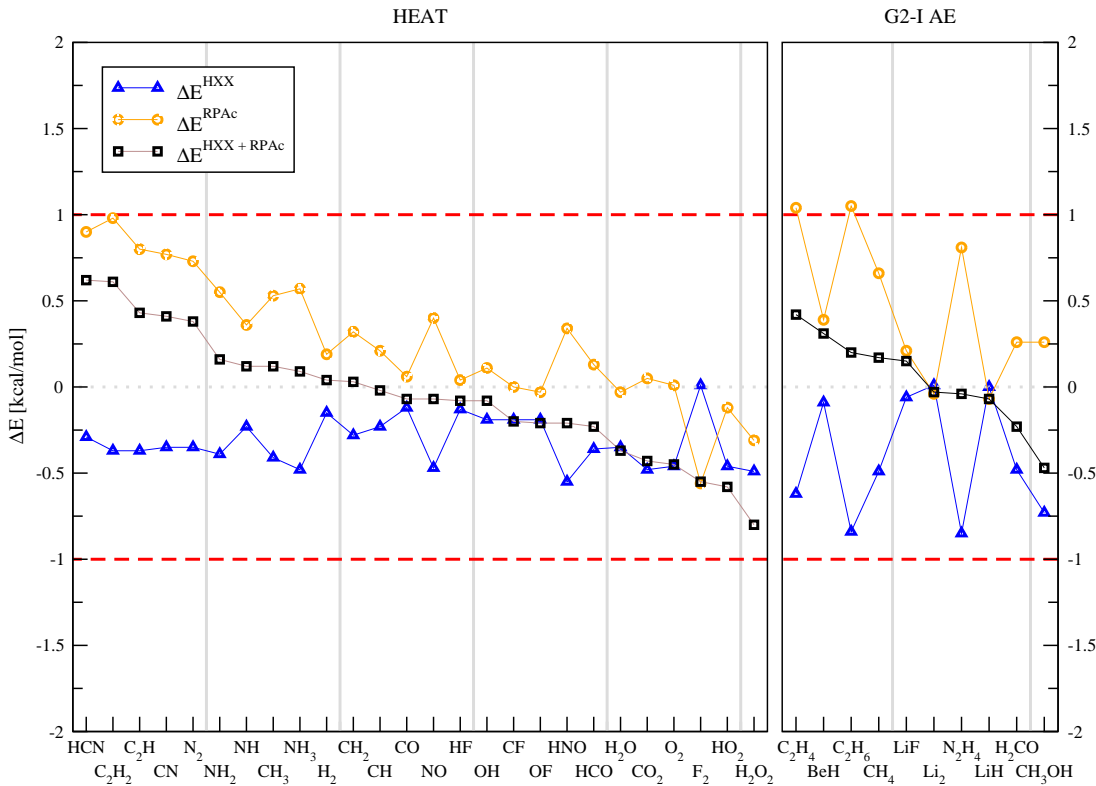}
    \caption{HXX (triangles), RPA correlation energy (circles) and total RPA (squares) atomization energy deviations between the plane-wave and GTO results (Table \ref{tab.4.1} and \ref{tab.4.2} respectively) for the HEAT set (left panel) and selected molecules of the G2-1 test set (right panel). Energy differences are given in \si{\kilo cal\per\mol}. The molecules are sorted in decreasing order by their deviation in the RPA atomization energy. The dashed lines mark the threshold of chemical precision (\SI{1}{\kilo cal\per\mol}).}
    \label{fig.4.1}
\end{figure*}

Fig.~\ref{fig.4.1} shows the deviation of the plane-wave atomization energies from the GTO calculations for the HXX and total RPA atomization energies as well as the RPA correlation energy (RPAc) contribution to the atomization energy. We observe that the plane-wave RPAc energies tend to over-correlate in comparison to the GTO results. Nevertheless, the reported RPAc energies are within the desired \SI{1}{\kilo cal \per\mol} accuracy--except for the molecules C$_2$H$_4$ and C$_2$H$_6$ whose correlation energies deviate by 1.04 and \SI{1.05}{\kilo cal\per\mol} respectively. 
In contrast to the HXX atomization energies, we observe that the RPAc energies are systematically less biased. 

Furthermore, Fig.~\ref{fig.4.1} shows that there is some degree of error cancellation between the exchange and correlation energy contributions. One notes that the dips in the yellow line ($\Delta E^{\rm RPAc}$) are mirrored by corresponding dips with an alternate sign in the blue line ($\Delta E^{\rm HXX}$). Thus, whenever the exchange energy is more strongly underestimated, the correlation energy RPAc is more strongly overestimated. The error in the combined exchange-correlation energy is therefore on average reduced compared to the individual errors in HXX and RPAc. On average there is no clear tendency for over- nor underbinding in the PAW method for the final atomization energies, which is also supported by the comparatively small mean signed deviation (MSD) in Table \ref{tab.4.2}.

Of particular interest are again those cases where we treat all electrons exactly in the PAW case: H$_2$, Li$_2$, LiH and BeH. It is clear that for these molecules, both the HXX and RPAc errors are small, and agreement for GTO and PAW is well within the combined estimated error bars of 0.3 kcal/mol. As to why there is some error cancellation we can only speculate: it is clear that the RPA screens the exchange interaction between the electrons, that is, in essence the bare exchange is screened by an electron cloud related to the polarizability of the electron gas. This might imply that errors in the description of the bare exchange are also partly canceled (screened) by the response of the electron gas. 
This compensation is, however, not exact; a residual error is clearly observed. Note that we have ordered the molecules according to their residual difference between plane waves and GTOs, but inspection of the order of the molecules does not suggest any systematics for this residual error. Possible sources are
i) the fact that we keep the $1s$ orbital frozen at the level of the PBE atoms in the PAW case, whereas they are fully relaxed using the PBE functional for the GTO case, ii) residual shape approximations used when evaluating the PAW RPAc correlation energy, iii) prevailing technical errors such as the extrapolation to infinite unit cell volume and basis-set extrapolation errors in the plane-wave calculations, but also remaining small basis-set truncation errors for the GTO case. The latter two errors might well still amount to 0.2--0.3 kcal/mol. To be certain that some of the prevailing errors are also related to i) and ii), we spot-checked one of the outliers, H$_2$O$_2$, very carefully. For this specific case, increasing the plane-wave energy cutoff or the box size does not significantly improve the agreement between the PAW and GTO methods. Also, different PAW potentials yield consistent results within 0.2 kcal/mol. We conclude that the VASP PAW results can hardly be improved for H$_2$O$_2$ and that some of the remaining errors are likely related to the PAW technique, either the freezing of the core orbitals at the level of the oxygen atom in the PBE, or other small numerical inaccuracies introduced during the fairly complex construction of the PAW potentials.

To investigate the impact of including the $1s$ core states in the correlated calculations, we also evaluated the GTO RPA correlation energies when explicitly correlating the $1s$ orbitals for the C, N, O, and F atoms. The comparison of the all-electron GTO calculations to the plane-wave results (still treating the $1s$ electrons as core states) is provided in the Supplementary material. We find that including the core states in the GTO calculations leads to a sizable increase of the atomization energies of \SI{0.8}{\kilo cal\per\mol} per C atom, \SI{0.4}{\kilo cal\per\mol} per N atom, \SI{0.45}{\kilo cal\per\mol} per O atom and \SI{0.3}{\kilo cal\per\mol} per F atom. Furthermore, the atomization energies slightly increase with the number of bonds in the molecule. Clearly, this residual error is larger than the differences between the PAW and GTO calculations when the $1s$ core states are consistently uncorrelated. 

We conclude that some of the residual PAW errors in the HXX and RPAc energies compensate each other, resulting in a very good agreement of the total RPA atomization energies. While BSSE and finite-size errors (volume and plane-wave basis-set size errors) hardly contribute to the discrepancies observed for the HXX energies, this can not be entirely ruled out for the RPA correlation energies. The observed discrepancy might originate from an interplay of the BSSE, errors in the extrapolation to infinite volume and infinite basis-set size, as well as residual PAW errors. Nevertheless, basis-set extrapolation and unit-cell-size extrapolation are necessary in order to obtain accurate RPA atomization energies. If done carefully, both approaches agree well within the desired accuracy of \SI{1}{\kilo cal\per\mol} for all investigated molecules. This confirms that consistent RPA correlation energies between different basis sets can be obtained by a careful convergence 
study and suitable PAW choices in the plane-wave calculation.

\section{Conclusion}\label{V}

The present work is motivated by the need of validation of PAW potentials for correlated wavefunction calculations. PAW potentials are  usually constructed for KS ground-state calculations and their accuracy is rarely assessed for such correlated calculations. We decided to evaluate the potentials for the atomization energies of molecules, since comparison to high-quality GTO calculations is at least in principle straightforward. The random phase approximation was chosen as a "simple" but prototypical correlated method, since its favourable scaling allows to achieve accurate and technically  converged results using both types of basis sets, but the method also exposes many of the problems present in more involved methods. To this end, reference type RPA atomization energies for the HEAT set as well as $10$ selected molecules of the G2-1 set were calculated using plane waves and the PAW method, and GTOs. 
All calculations were performed on top of PBE ground-state calculations, using PBE orbitals and one-electron energies.

After reviewing the basic theory, we discussed the technical details and strategies required to obtain highly accurate RPA atomization energies using plane waves in the PAW method as well as using GTOs. Regarding the plane-wave calculations, special care must be devoted to the treatment of the $\bm{G} = 0$ component in the exchange and correlation energy. Furthermore, while obtaining cell and basis-set converged HXX energies is straightforward by increasing the cell size and plane-wave cutoffs, this is computationally still rather challenging for the RPA correlation energy. Convergence of the RPA correlation energy with respect to the plane-wave basis-set size was achieved by an extrapolation with respect to the dimension of the response matrix. Furthermore, long-range vdW-like interactions between repeated images were removed via a cell size extrapolation. This was found to be particularly important for systems with spatially diffuse molecular orbitals, such as the considered Li-containing molecules and BeH. Likewise, the Gaussian basis-set correlation energies were evaluated at the complete basis-set limit using an explicitly correlated dRPA-F12 method. We expect the uncertainties of both methods with respect to the complete-basis-set limit to be smaller than \SI{0.2}{\kilo cal\per\mol}. Using these setups, we compared the plane-wave and GTO results. The two methods are found to agree within chemical accuracy (\SI{1}{\kilo cal\per\mol}) for the atomization energies of all considered molecules. We report root-mean-square deviations of \SI{0.41} and \SI{0.33}{\kilo cal\per\mol} for the HXX and RPA energies, respectively.

For the HXX atomization energies, we observed a systematic underbinding using the plane-wave results. Since we do not expect BSSE or finite-size errors to be present in these calculations, we attribute this difference to the frozen core approximation employed in the PAW method. More specifically, the $1s$ core states of the atoms (except for H, Li, and Be) were kept frozen at the level of the PBE reference atom calculations and were not relaxed in the preparatory PBE mean-field calculations. This means that our plane-wave HXX energies are upper bounds, and relaxation effects  that are expected to be larger for the molecules will increase the binding energies in the plane-wave calculations potentially bringing them in line with the GTO HXX results. 

For the RPA correlation energy, we found a reasonably systematic trend for over-correlation, but the relative error is not as consistent among different molecules as it was for HXX. We gather that besides a residual PAW error and the just discussed froze core error, also BSSE and extrapolation errors contribute to the discrepancies between the two codes (the combined error is likely to be 0.3 kcal/mol). Nevertheless, the opposing systematic trends for the HXX and RPA correlation energies results in an error cancellation leading to the very satisfactory agreement of the final RPA atomization energies (RMSD 0.33 kcal/mol), with differences being below the threshold of chemical accuracy.

In summary, the present work reports reference-type RPA atomization energies for the HEAT set and selected molecules of the G2-1 set for GTOs and plane waves. Our assessment convincingly demonstrates that the employed PAW potentials are applicable to correlated wavefunction calculations and that the residual error due to the PAW approximation is below the threshold of chemical accuracy. Going forward, this validates the use of the PAW potentials in solid-state correlated wavefunction calculations where reference calculations are not yet easily and systematically doable.

\section*{Supplementary Material}
See the supplementary material for scalar relativistic HXX and RPA atomization energies using plane-wave and Gaussian basis sets as well as the comparison of all-electron GTO RPA atomization energies to the plane-wave results.

\section*{Acknowledgment}
Funding by the Austrian Science foundation (FWF) within the project  P 33440 is gratefully acknowledged. MEH and WK acknowledge support by the German Bundesministerium für Bildung und Forschung (BMBF) through the Helmholtz research program ``Materials Systems Engineering'' (MSE).

\section*{Author declarations}
\subsection*{Conflict of Interest}
The authors have no conflicts to disclose.
\section*{Data availability}
The data that support the findings of this study are available within the article and its supplementary material.

\nocite{*}
\bibliography{bibliography.bib}

\end{document}


\title{Supporting information for: \linebreak
``Approaching the basis-set limit of the dRPA correlation energy with explicitly correlated and projector augmented wave methods''}

\author{Moritz Humer}
\affiliation{University of Vienna, Faculty of Physics and Center for Computational Materials Science, Kolingasse 14-16, A-1090 Vienna, Austria}
\affiliation{University of Vienna, Faculty of Physics \& Vienna Doctoral School in Physics,  Boltzmanngasse 5, A-1090 Vienna, Austria}
\author{Michael E. Harding}
\affiliation{Institut f\"{u}r Nanotechnologie, Karlsruher Institut f\"u{r} Technologie (KIT), Campus Nord, Postfach 3640, D-76021 Karlsruhe, Germany}
\author{Martin Schlipf}
\affiliation{VASP Software GmbH, Sensengasse 8, 1090 Vienna, Austria}
\author{Amir Taheridehkordi}
\affiliation{University of Vienna, Faculty of Physics and Center for Computational Materials Science, Kolingasse 14-16, A-1090 Vienna, Austria}
\author{Zoran Sukurma}
\affiliation{University of Vienna, Faculty of Physics and Center for Computational Materials Science, Kolingasse 14-16, A-1090 Vienna, Austria}
\affiliation{University of Vienna, Faculty of Physics \& Vienna Doctoral School in Physics,  Boltzmanngasse 5, A-1090 Vienna, Austria}
\author{Wim Klopper}
\affiliation{Institut f\"{u}r Nanotechnologie, Karlsruher Institut f\"u{r} Technologie (KIT), Campus Nord, Postfach 3640, D-76021 Karlsruhe, Germany}
\affiliation{Institut f\"{u}r Physikalische Chemie, Karlsruher Institut f\"u{r} Technologie (KIT), Campus S\"{u}d, Postfach 6980, D-76049 Karlsruhe, Germany}
\author{Georg Kresse}
\affiliation{University of Vienna, Faculty of Physics and Center for Computational Materials Science, Kolingasse 14-16, A-1090 Vienna, Austria}
\affiliation{VASP Software GmbH, Sensengasse 8, 1090 Vienna, Austria}

\date{\today}
\maketitle
\pagebreak

\setlength{\LTcapwidth}{0.75\textwidth}
\begin{longtable}{@{\extracolsep{\fill}}lrrr}
\caption{\label{tab.5.1}Hartree plus exact exchange (HXX) atomization energies using PBE orbitals for the HEAT test set and selected molecules of the G2-1 test set.
The energies were calculated using plane-wave (PW) and Gaussian basis sets (GTO) including scalar relativistic corrections. Energies 
are given in kcal/mol. The differences between PW and GTO are displayed in the column $\Delta_{\rm PW - GTO}$. Additionally, the root-mean-square deviation (RMSD) and the mean signed deviation (MSD) are given in the final rows.} \\
\hline\hline
Molecule & \multicolumn{1}{c}{PW} & \multicolumn{1}{c}{GTO} & \multicolumn{1}{c}{$\Delta_{\rm PW - GTO}$} \\ \hline
\endfirsthead
\hline\hline
Molecule & \multicolumn{1}{c}{PW} & \multicolumn{1}{c}{GTO} & \multicolumn{1}{c}{$\Delta_{\rm PW - GTO}$} \\ \hline
\endhead
\hline
\hline
\endfoot
C$_2$H     & 175.94                   & 176.18                        & -0.24                                              \\
C$_2$H$_2$ & 290.23                   & 290.38                        & -0.15                                              \\
CF         & 72.06                    & 72.27                         & -0.21                                              \\
CH         & 55.86                    & 56.00                         & -0.14                                              \\
CH$_2$     & 154.07                   & 154.17                        & -0.10                                              \\
CH$_3$     & 242.22                   & 242.34                        & -0.12                                              \\
CN         & 71.00                    & 71.35                         & -0.35                                              \\
CO         & 169.27                   & 169.40                        & -0.13                                              \\
CO$_2$     & 233.22                   & 233.70                        & -0.48                                              \\
F$_2$      & -42.77                   & -42.73                        & -0.04                                              \\
H$_2$      & 84.08                    & 84.07                         & 0.01                                               \\
H$_2$O     & 154.58                   & 154.85                        & -0.27                                              \\
H$_2$O$_2$ & 128.51                   & 128.95                        & -0.44                                              \\
HCN        & 194.36                   & 194.55                        & -0.19                                              \\
HCO        & 175.23                   & 175.51                        & -0.28                                              \\
HF         & 96.04                    & 96.17                         & -0.13                                              \\
HNO        & 71.61                    & 72.10                         & -0.49                                              \\
HO$_2$     & 58.39                    & 58.84                         & -0.45                                              \\
N$_2$      & 110.40                   & 110.76                        & -0.36                                              \\
NH         & 49.74                    & 49.90                         & -0.16                                              \\
NH$_2$     & 115.98                   & 116.22                        & -0.24                                              \\
NH$_3$     & 199.53                   & 199.79                        & -0.26                                              \\
NO         & 46.11                    & 46.60                         & -0.49                                              \\
O$_2$      & 25.13                    & 25.61                         & -0.48                                              \\
OF         & -30.79                   & -30.57                        & -0.22                                              \\
OH         & 67.28                    & 67.42                         & -0.14                                              \\
BeH        & 49.24                    & 49.25                         & -0.01                                              \\
C$_2$H$_4$ & 424.40                   & 424.63                        & -0.23                                              \\
C$_2$H$_6$ & 547.68                   & 547.96                        & -0.28                                              \\
H$_2$CO    & 250.87                   & 251.16                        & -0.29                                              \\
CH$_4$     & 326.88                   & 327.00                        & -0.12                                              \\
CH$_3$OH    & 364.11                   & 364.51                        & -0.40                                              \\
Li$_2$     & 3.49                     & 3.48                          & 0.01                                               \\
LiH        & 33.87                    & 33.80                         & 0.07                                               \\
LiF        & 86.75                    & 86.81                         & -0.06                                              \\
N$_2$H$_4$ & 262.49                   & 263.04                        & -0.55 
\\
\hline
RMSD    & & & 0.28
\\
MSD     & & & -0.23
\end{longtable}

\setlength{\LTcapwidth}{0.745\textwidth}
\begin{longtable}{@{\extracolsep{\fill}}lrrr}
\caption{\label{tab.5.2}RPA atomization energies using PBE orbitals for the HEAT test set 
and selected molecules of the G2-1 test set. The energies were calculated 
using plane-wave (PW) and Gaussian basis sets (GTO)
using scalar relativistic corrections. Energies are given in kcal/mol. The differences between PW and GTO are displayed in the column $\Delta_{\rm PW - GTO}$. Additionally, the root-mean-square deviation (RMSD) and the mean signed deviation (MSD) are given in the final rows.} \\
\hline\hline
Molecule & \multicolumn{1}{c}{PW} & \multicolumn{1}{c}{GTO} & \multicolumn{1}{c}{$\Delta_{\rm PW - GTO}$} \\ \hline
\endfirsthead
\hline\hline
Molecule & \multicolumn{1}{c}{PW} & \multicolumn{1}{c}{GTO} & \multicolumn{1}{c}{$\Delta_{\rm PW - GTO}$} \\ \hline
\endhead
\hline
\hline
\endfoot
C$_2$H     & 244.82                   & 244.27                        & 0.55                                               \\
C$_2$H$_2$ & 381.72                   & 380.91                        & 0.81                                               \\
CF         & 120.13                   & 120.43                        & -0.30                                              \\
CH         & 81.23                    & 81.17                         & 0.06                                               \\
CH$_2$     & 179.65                   & 179.45                        & 0.20                                               \\
CH$_3$     & 294.60                   & 294.23                        & 0.37                                               \\
CN         & 172.60                   & 172.17                        & 0.43                                               \\
CO         & 244.25                   & 244.31                        & -0.06                                              \\
CO$_2$     & 363.90                   & 364.31                        & -0.41                                              \\
F$_2$      & 29.84                    & 30.54                         & -0.70                                              \\
H$_2$      & 108.86                   & 108.69                        & 0.17                                               \\
H$_2$O     & 222.64                   & 222.93                        & -0.29                                              \\
H$_2$O$_2$ & 254.72                   & 255.46                        & -0.74                                              \\
HCN        & 299.37                   & 298.66                        & 0.71                                               \\
HCO        & 263.25                   & 263.39                        & -0.14                                              \\
HF         & 132.25                   & 132.40                        & -0.15                                              \\
HNO        & 198.72                   & 198.87                        & -0.15                                              \\
HO$_2$     & 165.20                   & 165.75                        & -0.55                                              \\
N$_2$      & 223.58                   & 223.21                        & 0.37                                               \\
NH         & 82.47                    & 82.28                         & 0.19                                               \\
NH$_2$     & 179.30                   & 179.02                        & 0.28                                               \\
NH$_3$     & 290.59                   & 290.31                        & 0.28                                               \\
NO         & 147.66                   & 147.75                        & -0.09                                              \\
O$_2$      & 112.67                   & 113.13                        & -0.46                                              \\
OF         & 45.99                    & 46.33                         & -0.34                                              \\
OH         & 103.23                   & 103.28                        & -0.05                                              \\
BeH        & 50.92                    & 50.55                         & 0.37                                               \\
C$_2$H$_4$ & 537.76                   & 536.98                        & 0.78                                               \\
C$_2$H$_6$ & 684.02                   & 683.31                        & 0.71                                               \\
H$_2$CO    & 355.33                   & 355.37                        & -0.04                                              \\
CH$_4$     & 405.09                   & 404.59                        & 0.50                                               \\
CH$_3$OH   & 490.22                   & 490.39                        & -0.17                                              \\
Li$_2$     & 18.89                    & 18.91                         & -0.02                                              \\
LiH        & 54.45                    & 54.48                         & -0.03                                              \\
LiF        & 127.14                   & 126.99                        & 0.15                                               \\
N$_2$H$_4$ & 426.35                   & 426.13                        & 0.22                                               \\
\hline
RMSD    & & & 0.40
\\
MSD     & & & 0.07
\end{longtable}

\setlength{\LTcapwidth}{0.745\textwidth}
\begin{longtable}{@{\extracolsep{\fill}}lrrr}
\caption{\label{tab.5.3}All-electron RPA atomization energies using PBE orbitals for the HEAT 
test set and selected molecules of the G2-1 test set. The energies 
were calculated using Gaussian basis sets (GTO) without relativistic 
corrections and are compared to the plane-wave results (PW). Energies are given in kcal/mol. Energies are given in kcal/mol. The differences between PW and GTO are displayed in the column $\Delta_{\rm PW - GTO}$. Additionally, the root-mean-square deviation (RMSD) and the mean signed deviation (MSD) are given in the final rows.} \\
\hline\hline
Molecule & \multicolumn{1}{c}{PW} & \multicolumn{1}{c}{GTO} & \multicolumn{1}{c}{$\Delta_{\rm PW - GTO}$} \\ \hline
\endfirsthead
\hline\hline
Molecule & \multicolumn{1}{c}{PW} & \multicolumn{1}{c}{GTO} & \multicolumn{1}{c}{$\Delta_{\rm PW - GTO}$} \\ \hline
\endhead
\hline
\hline
\endfoot
C$_2$H     & 244.95                   & 246.60                        & -1.65                                              \\
C$_2$H$_2$ & 381.78                   & 383.62                        & -1.84                                              \\
CF         & 120.38                   & 121.04                        & -0.66                                              \\
CH         & 81.19                    & 81.40                         & -0.21                                              \\
CH$_2$     & 179.62                   & 180.33                        & -0.71                                              \\
CH$_3$     & 294.51                   & 295.39                        & -0.88                                              \\
CN         & 172.73                   & 173.62                        & -0.89                                              \\
CO         & 244.39                   & 245.64                        & -1.25                                              \\
CO$_2$     & 364.33                   & 366.85                        & -2.52                                              \\
F$_2$      & 30.01                    & 30.61                         & -0.60                                              \\
H$_2$      & 108.73                   & 108.69                        & 0.04                                               \\
H$_2$O     & 222.82                   & 223.60                        & -0.78                                              \\
H$_2$O$_2$ & 255.01                   & 256.34                        & -1.33                                              \\
HCN        & 299.49                   & 300.70                        & -1.21                                              \\
HCO        & 263.42                   & 264.94                        & -1.52                                              \\
HF         & 132.51                   & 132.78                        & -0.27                                              \\
HNO        & 198.93                   & 199.80                        & -0.87                                              \\
HO$_2$     & 165.42                   & 166.43                        & -1.01                                              \\
N$_2$      & 223.72                   & 224.48                        & -0.76                                              \\
NH         & 82.47                    & 82.52                         & -0.05                                              \\
NH$_2$     & 179.33                   & 179.57                        & -0.24                                              \\
NH$_3$     & 290.64                   & 291.25                        & -0.61                                              \\
NO         & 147.86                   & 148.63                        & -0.77                                              \\
O$_2$      & 112.85                   & 113.74                        & -0.89                                              \\
OF         & 46.19                    & 46.53                         & -0.34                                              \\
OH         & 103.31                   & 103.58                        & -0.27                                              \\
BeH        & 50.88                    & 50.57                         & 0.31                                               \\
C$_2$H$_4$ & 537.72                   & 539.60                        & -1.88                                              \\
C$_2$H$_6$ & 683.88                   & 685.92                        & -2.04                                              \\
H$_2$CO    & 355.46                   & 357.08                        & -1.62                                              \\
CH$_4$     & 404.94                   & 405.94                        & -1.00                                              \\
CH$_3$OH   & 490.36                   & 492.23                        & -1.87                                              \\
Li$_2$     & 18.88                    & 18.91                         & -0.03                                              \\
LiH        & 54.41                    & 54.48                         & -0.07                                              \\
LiF        & 127.35                   & 127.36                        & -0.01                                              \\
N$_2$H$_4$ & 426.59                   & 427.82                        & -1.23                                                                                          \\
\hline
RMSD    & & & 1.10
\\
MSD     & & & -0.88
\end{longtable}